\documentclass[aps,prb,floatfix,reprint]{revtex4-1}
\usepackage{amsmath}
\usepackage{amsfonts}
\usepackage{amssymb}
\usepackage{graphics}
\usepackage{float}
\usepackage{graphicx}
\usepackage{epstopdf}
\usepackage[compact]{titlesec}
\usepackage{natbib}
\usepackage{threeparttable}
\usepackage[titletoc,title]{appendix}
\usepackage{lipsum}
\usepackage{varwidth}
\usepackage{tabularx}
\usepackage{color}
\newcommand{\myquad}[1][1]{\hspace*{#1em}\ignorespaces}

\begin{document}
\title{Anomalous heat flow in 8-$Pmmn$ borophene with tilted Dirac cones}
\author{Parijat Sengupta}
\affiliation{Dept. of Electrical Engineering, University of Illinois, Chicago, IL 60607 USA}
\author{Yaohua Tan}
\affiliation{Dept. of Electrical Engineering, University of Virginia, Charlottesville, VA 22904 USA}
\author{Enrico Bellotti}
\affiliation{Dept. of Electrical Engineering, Boston University, Boston, MA 02215 USA}
\author{Junxia Shi}
\affiliation{Dept. of Electrical Engineering, University of Illinois, Chicago, IL 60607 USA}

\begin{abstract}
We analytically establish an anomalous transverse flow of heat in 8-\textit{Pmmn} borophene, one of the several two-dimensional (2D) allotropes of Boron (B). The dispersion of this allotrope contains a pair of anisotropic and tilted Dirac cones which are gapped by placing the 2D \textit{B} sheet under an intense circularly-polarized illumination. A gap in the Dirac dispersion leads to a finite Berry curvature and connected anomalous Hall effects. In the case of thermoelectrics, this manifests as a heat current perpendicular to the temperature gradient - the thermal Hall effect. A quantitative calculation of the attendant thermal Hall conductivity reveals dependence on the intrinsic anisotropy and tilt of the Dirac cone. Further, by estimating the longitudinal thermal conductivity using the Weidemann-Franz law, we also outline steps to compute the thermal Hall angle that gauges the generation efficiency of such transverse heat processes. Finally, we touch upon the idea of thermal rectification wherein the direction of flow of the anomalous heat reverses through a simple switch of the polarization of incident light and is of interest in thermal logic circuits.  
\end{abstract}
\maketitle

\vspace{0.3cm}
\section{Introduction}
\vspace{0.3cm}
The two-dimensional (2D) boron monolayers known as borophenes~\cite{mannix2017synthesis,kong2018recent,zhou2016low} and have been proposed and synthesized in a variety of allotropes. Prominent examples of such 2D boron sheets predicted through first-principles calculation and eventually grown on silver (Ag) substrates are the striped, $ \beta_{12} $, and $\chi_{3}$ allotropes. Boron, in general, exhibits multiple phases leading to the formation of structures such as quasi-planar clusters, cage-like fullerenes, and nanotubes. The existence of multiple allotropes is explained by taking recourse to the electron-deficient character of the atom - it possesses three valence electrons and four orbitals - from which emerges several complex bonding patterns to fulfill requirements of crystal stability. In parallel with development of laboratory techniques~\cite{mannix2015synthesis} to realize stable-variants of borophenes, a large body of first-principles study~\cite{zhou2014semimetallic} employing structural search algorithms predicted the possibility of highly stable 2D boron sheets of finite thickness. In particular, a free-standing arrangement of two-dimensional boron atoms with a buckled structure (Fig.~\ref{ubcell}) and an orthorhombic 8-$ Pmmn $ symmetry ($ Pmmn $ represents the space group 59; 8 denotes the number of atoms in the unit cell) was shown to carry anisotropic and tilted Dirac cones (DCs); it has since been theoretically expanded.~\cite{lopez2016electronic}

In this letter, we examine anomalous thermal transport when the 8-$ Pmmn $ 2D boron allotrope (henceforth, referred to as $\beta $-borophene) with DCs is irradiated by a strong optical beam. Before elaborating on the `anomalous' part, we set the context by noting that $\beta $-borophene with tilted DCs, a new Dirac material in the family of 2D layered structures, holds promise as a thermoelectric material~\cite{carrete2016physically} with a thermal conductivity matching monolayer MoS$_{2}$ and carrying an innate anisotropy, the very feature that endows 2D layered black phosphorus to possess a very high thermoelectric figure-of-merit. In addition to `normal' heat flow, an additional contribution may arise from the non-trivial band topology of $ \beta $-borophene; the DCs when gapped carry a finite Berry curvature $\left(\Omega\left(k\right)\right)$ helping establish a class of phenomena collectively recognized as anomalous Hall effect $ \left(AHE\right) $ observable in charge, spin, and heat currents.~\cite{xiao2010berry} The $ \beta $-borophene is intrinsically gapless and thus precludes the occurrence of $ \Omega\left(k\right) $ and the aforementioned set of Hall effects. It is however now well-understood and verified~\cite{wang2013observation} that in pristine Dirac materials (zero gap) strain-engineering or a suitable choice of substrates can lead to a gapped phase; additionally, photo-irradiation is also known to rearrange electronic states and in a special case known as the \textit{off}-resonant approximation, a band-gap $\left(\Delta\right)$ opening can be achieved. Here, under the \textit{off}-resonant condition, we estimate a photo-induced $ \Delta $ in $ \beta $-borophene and compute the resulting non-zero $ \Omega\left(k\right) $; in conjunction with a temperature gradient $\left(\nabla T\right)$, the $ \Omega\left(k\right) $ sets up a heat current transverse to vector of $ \nabla T $.  A similar heat flow is observed~\cite{abrikosov2017fundamentals} when a real-space magnetic field is applied instead of $ \Omega\left(k\right) $ - the well-known Rigi-Leduc effect ($RLE$). 

As a brief summary of the results presented herein, we note that the thermal Hall conductance rises for a lower photo-induced $ \Delta $ and is further enhanced for a Fermi level located in the band gap. In addition, the conductance also shows a monotonic rise (fall) as the anisotropy of $ \beta $-borophene is augmented (truncated) via a strain-like perturbation. In fact, we outline how both these observations are encapsulated in a general expression of the thermal Hall angle that potentially estimates the efficiency of the $ \Omega\left(k\right) $-aided anomalous heat flow. Lastly, the polarization of the \textit{off}-resonant light beam when switched between a right- or left-circular type allows for a full reversal of the direction of flow of the anomalous heat current and espouses the principle of thermal rectification. 

\vspace{0.3cm}
\section{Theory and Model}
\begin{figure}[t!]
\includegraphics[scale=0.52]{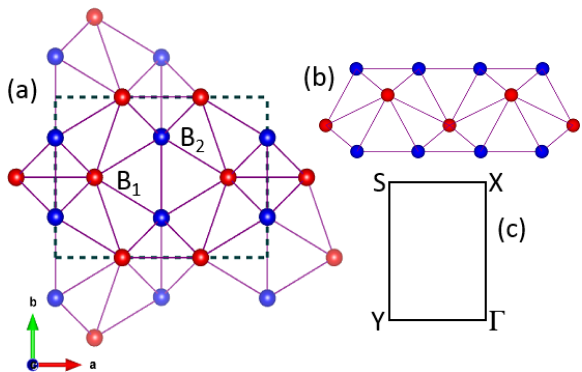}
\caption{The eight-atom simple orthorhombic arrangement in the $ \beta $-borophene unit cell (marked by dotted lines) is illustrated here. The in-plane lattice constants are  $ a = 0.452\,nm $ and $ b = 0.326\,nm $ while $ c = 1.3\,nm $ is out-of-plane. There are two inequivalent atom positions indicated by the identifiers, B$_{1}$ and B$_{2}$, in sub-figure (a). The coordinates of B$_{1}$ are $\left(0.185a, b/2, 0.531c\right)$ while those for B$_{2}$ can be written as $\left(a/2, 0.247b,0.584c\right)$; the difference in the \textit{z}-coordinate represents the intrinsic buckling (b). The positions for the remaining six atoms follow from the symmetry operations of space-group 59. The bonds between the boron atoms are denoted by solid lines. The high-symmetry points in the corresponding Brillouin zone are shown in (c); the tilted Dirac cone in $\beta $-borophene is formed along the $ \Gamma - Y $ direction.}
\label{ubcell}
\vspace{-0.5cm}
\end{figure}

\begin{figure}
\includegraphics[scale=0.52]{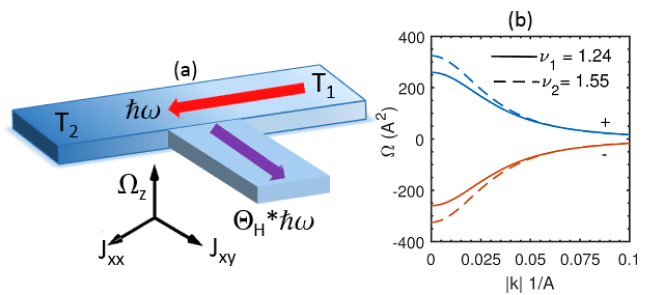}
\vspace{-0.28cm}
\caption{A generalized illustration of the thermal Hall effect caused by $\left(\Omega\left(k\right)\right)$ directed out-of-plane (for $\Delta = 0.2\,eV $) is presented in the left figure (a). The anomalous flow of transverse heat current $\left(\hbar\omega\right)$ in presence of a temperature gradient $\left(T_{1} > T_{2}\right)$ is gauged via the thermal Hall conductivity. For the case of $ \beta $-borophene, the $\Omega\left(k\right)$ curves plotted in the right panel (b) using Eq.~\ref{bc} is photo-induced (see Eq.~\ref{commsf}). The set of upper (lower) curves collectively identified by `+' (`-') are for the conduction (valence) band and are shown here for a pair of $ \nu = v_{x}/v_{y} $ values indicated on the plot. Note that $ \kappa = 1.246 $ is the intrinsic anisotropy of $ \beta $-borophene while external perturbation, for example, strain can induce additional distortion. For an artificially assigned higher $ \kappa $ obtained by reducing the $ v_{y} $ component, $\Omega\left(k\right)$ can be tuned thus offering a possible way to manipulate the thermal Hall angle, $\Theta_{H}$, in addition to the photo-dependent scheme.}
\vspace{-0.4cm}
\label{berth}
\end{figure}

\vspace{0.3cm}
To expand on these findings, we begin by writing the low-energy continuum two-band Hamiltonian~\cite{zabolotskiy2016strain} for $ \beta $-borophene that describes an anisotropic and tilted Dirac crossing along the $ \Gamma $-Y direction in the Brillouin zone (see Fig.~\ref{ubcell}(c)).
\begin{equation}
H = \hbar v_{x}\sigma_{x}p_{x} + \hbar v_{y}\sigma_{y}p_{y} + \hbar v_{t}\sigma_{0}p_{y}.
\label{boeqn}
\end{equation}
In Eq.~\ref{boeqn}, $ \sigma_{x,y} $ are the Pauli matrices denoting the lattice degree of freedom while $ \sigma_{0} $ is  the $ 2 \times 2 $ identity matrix. The direction-dependent velocity terms $\left(\times 10^{6}\,m/s \right)$ as reported in Ref.~\onlinecite{zabolotskiy2016strain} are $ v_{x} = 0.86 $, $ v_{y} = 0.69 $, and $ v_{t} = 0.32 $. Note that anisotropy, $ \nu = v_{x}/v_{y} $, arises since $ \nu \neq 1 $ while $ v_{t} \neq 0 $ ensures a tilt through non-concentric constant energy contours. The dispersion relation using Eq.~\ref{boeqn} is
\begin{equation}
E_{\pm}\left(k\right) = \hbar k\left(v_{t}\sin\left(\theta\right) \pm\sqrt{v_{x}^{2}\cos^{2}\theta + v_{y}^{2}\sin^{2}\theta}\right).
\label{diseq}
\end{equation}
The upper (lower) sign is for the conduction (valence) band in $\beta $-borophene. This basic energy description (Eqs.~\ref{boeqn},~\ref{diseq}) serves as the starting point to examine models of light-matter interaction. 
The light source is an external electromagnetic perturbation periodic in time and serves to gap the Dirac dispersion. To quantitatively assess the change in dispersion under a periodic (with time period $ \mathcal{T} $) perturbation, we must make use of the Floquet theory~\cite{cayssol2013floquet} for a periodic Hamiltonian, $ \hat{H}\left(t\right) = \hat{H}\left(t + \mathcal{T}\right) $ to construct the corresponding eigen states. Note that the time-dependent Hamiltonian is $ \hat{H}\left(t\right) = \hat{H}_{0} + \hat{V}\left(t\right)$, where $ \hat{H}_{0} $ is the stationary part and $ \hat{V}\left(t\right) $ is the time-periodic perturbation. The time-dependent Hamiltonian can be transformed into a Floquet Hamiltonian, which, following Tannor~\cite{tannor2007introduction}, is $ \hat{H}_{F}\left(t\right) =  \left(\hat{H}\left(t\right) - i\hbar\partial_{t} \right) $. The Floquet Hamiltonian when solved takes a matrix form~\cite{tannor2007introduction} and recasts a time-dependent problem to a time-independent Schr{\"o}dinger equation. For cases where the light frequency $ \left(\omega\right) $ is much higher than the energy scales of the stationary Hamiltonian, an approximation to the Floquet Hamiltonian can be written~\cite{lopez2015photoinduced}
\begin{equation}
\hat{H}_{F}\left(k\right) = \hat{H}_{0} + \dfrac{\left[H_{-1}, H_{1}\right]}{\hbar\omega}.
\label{fleq1}
\end{equation}
The terms contained in the anti-commutator are the Fourier components which have the following generalized form: $ H_{m} = \dfrac{1}{\mathcal{T}}\int_{0}^{\mathcal{T}}dt \exp\left(im\omega t\right)H_{t} $. $ H_{t} $ is the time-dependent part of $ \hat{H}\left(t\right) $ and $ \mathcal{T} = 2\pi/\omega $. The Fourier components must be calculated using the Hamiltonian in Eq.~\ref{boeqn}. Changing into a time-dependent form by applying the Peierls substitution $ \hbar\mathbf{k} \rightarrow \hbar\mathbf{k} - e\mathbf{A}\left(t\right) $ that represents the coupling (via the vector potential $ \mathbf{A}\left(t\right) $) of the electromagnetic field, Eq.~\ref{boeqn} is
\begin{equation}
\begin{aligned}
\hat{H}\left(t\right) &=  v_{x}\sigma_{x}\left(\hbar k_{x} + eA_{x}\right) +  v_{y}\sigma_{y}\left(\hbar k_{y} + eA_{y}\right)  \\
& +  v_{t}\sigma_{0}\left(\hbar k_{y} + eA_{y}\right).
\label{fleq3}
\end{aligned}
\end{equation}
The time-dependent part is therefore, $ \hat{H}_{t} = v_{x}\sigma_{x}eA_{x} + v_{y}\sigma_{x}eA_{y} + v_{t}\sigma_{0}eA_{y} $. For circularly polarized light propagating along $ \hat{z} $, the two vector components are $ \mathbf{A} = A_{0}\left(s\sin\omega t\hat{x}, \cos\omega t\hat{y}\right) $. The right (left) circularly-polarized light is $ s = 1\left(-1\right) $. From the time-dependent part, the two Fourier components in Eq.~\ref{fleq1} $ \left(H_{\pm}\right) $ are
\begin{equation}
H_{\pm} = \dfrac{eA_{0}}{\mathcal{T}}\int_{0}^{\mathcal{T}}dte^{\pm i\omega t}\left[\phi_{1}\sin\omega t + \phi_{2}\cos\omega t\right],
\label{comms}
\end{equation}
where $ \phi_{1} = sv_{x}\sigma_{x} $ and $ \phi_{2} = \left(v_{t}\sigma_{0}+ v_{y}\sigma_{y}\right) $. Evaluating the integral in Eq.~\ref{comms} gives 
\begin{equation}
H_{\pm} = \dfrac{eA_{0}}{2}\left(\mp sv_{x}\sigma_{x} + v_{y}\sigma_{y} + v_{t}\sigma_{0}\right).
\label{intres}
\end{equation}
Inserting the Fourier components $ H_{\pm} $ from Eq.~\ref{intres} in the commutator of Eq.~\ref{fleq1} and using the Pauli relation, $ \left[\sigma_{x},\sigma_{y}\right] = 2i\sigma_{z} $, we obtain
\begin{equation}
\dfrac{\left[H_{-1}, H_{1}\right]}{\hbar\omega} = \Delta = \dfrac{se^{2}A_{0}^{2}v_{x}v_{y}\sigma_{z}}{\hbar\omega}.
\label{commsf}
\end{equation}
The quantity $ \Lambda = e^{2}A_{0}^{2}v_{x}v_{y} $ in Eq.~\ref{commsf} has unit of energy-squared and is an experimentally alterable parameter (in addition to $ \hbar\omega $) through adjustments to the experimental setup. Briefly, $ \Lambda $ denotes the power gained by an electron traveling with velocity $ v $ in an electric field $ E = A\omega $ in one period of the electromagnetic field of the illuminating beam. A simple inspection of Eqs.~\ref{fleq1} and~\ref{commsf} shows that the rearrangement of bands via the additional Floquet term introduced through Eq.~\ref{commsf} lets the original Hamiltonian in Eq.~\ref{boeqn} acquire a mass term, $ \Delta $. The gapping of the massless Dirac bands therefore manifests as a non-vanishing $ \Omega\left(k\right) $, in which lies the genesis of the family of $ AHE $. We therefore derive an analytic representation of $ \Omega\left(k\right) $ and subsequently evaluate the thermal variant of the $ AHE $ (see Fig.~\ref{berth}(a)).

For a general two-band model Hamiltonian of the form $ H = \mathbf{\sigma}\cdot \mathbf{d}\left(k\right) + \mathbb{I}\epsilon\left(k\right) $, where $ \mathbf{d} $ is a vector of spin or pseudo-spin, $ \epsilon\left(k\right) $ is a general dispersion term, and $ \mathbb{I} $ is the $ 2 \times 2 $ identity matrix, the Berry curvature for this system is~\cite{fruchart2013introduction} 
\begin{equation}
\Omega_{\mu\nu} = \dfrac{1}{2}\varepsilon_{\alpha\beta\gamma}\hat{d}_{\alpha}\left(\mathbf{k}\right)\partial_{k_{\mu}}\hat{d}_{\beta}\left(\mathbf{k}\right)\partial_{k_{\nu}}\hat{d}_{\gamma}\left(\mathbf{k}\right),
\label{berry}
\end{equation}
where $ \hat{\mathbf{d}}\left(\mathbf{k}\right) = \dfrac{\mathbf{d\left(\mathbf{k}\right)}}{d\left(k\right)} $. Applying this formalism to the $ \beta $-borophene Hamiltonian (Eq.~\ref{boeqn}), and noting that the vector $ \mathbf{d} $ in component notation assumes the form:
\begin{equation}
\mathbf{\hat{d}}\left(k\right) = \dfrac{\hbar}{\sqrt{\left(\hbar v_{x}k_{x}\right)^{2} + \left(\hbar v_{y}k_{y}\right)^{2} + \Delta^{2}}}\left(v_{x}k_{x},v_{y}k_{y}, \Delta\right).
\label{dmer}
\end{equation}
Substituting the $ \mathbf{d} $ vector in Eq.~\ref{berry}, $ \Omega\left(k\right) $ is expressed  as:
\begin{equation}
\Omega\left(k\right) = \mp\dfrac{\hbar^{2}v_{x}v_{y}\Delta}{2\left[\left(\hbar^{2}v_{x}^{2}k_{x}^{2} + \hbar^{2}v_{y}^{2}k_{y}^{2}\right)^2 + \Delta^2\right]^{3/2}}\hat{\mathbf{z}}.
\label{bc}
\end{equation}
The upper (lower) sign is for the conduction (valence) band. The $ \Omega\left(k\right) $ as a momentum-dependent magnetic field points out-of-plane (the \textit{z}-axis) and evidently decays as $ \Delta \rightarrow 0 $. A plot of $ \Omega\left(k\right) $ for $ \vert k \vert $ values centred around the Dirac crossing is shown in Fig.~\ref{berth}(b). The $ \Omega\left(k\right) $ as expected peaks around the $ \vert k \vert = 0 $ mark and diminishes as we move farther in momentum-space.

The preceding evaluation of $ \Omega\left(k\right) $ allows us to proceed with an analysis of the anomalous flow of heat current in $ \beta $-borophene; firstly we note that transport equations for charge $ \left(J_{e}\right) $ and heat $ \left(J_{q}\right) $ currents as a response to applied electric $ \left(-\nabla V \right) $ and temperature $ \left(-\nabla T\right) $ gradients are written as~\cite{goldsmid2010introduction}
\begin{equation}
\begin{pmatrix}
J_{e} \\
J_{q}
\end{pmatrix}  = \begin{pmatrix}
L_{11} & L_{12} \\
L_{21} & L_{22}
\end{pmatrix}\begin{pmatrix}
\mathbf{-\nabla V} \\
-\nabla T
\end{pmatrix}.
\label{bcheq}
\end{equation}
The transport coefficient of interest to us is the transverse component (note the superscripts) $ L_{22}^{xy} $ - a measure of heat current in the direction of the \textit{y}-axis when a temperature gradient exists along the longitudinal \textit{x}-axis and the setup is subjected to an out-of-plane magnetic field. In the present context, the part of an external magnetic field is fulfilled by $ \Omega\left(k\right) $. The transverse thermal transport coefficient $ L_{22}^{xy} $ is~\cite{bergman2010theory,matsumoto2011theoretical}
\begin{align}
L_{22}^{xy} &= \dfrac{k^{2}_{B}T}{4\pi^{2}\hbar}\int \mathbf{d}^{2}k \Omega\left(k\right)\biggl[\dfrac{\pi^{2}}{3}-\left(ln\left(1-f\left(E\right)\right)\right)^{2}\notag \\ 
& \myquad[-2]-2Li_{2}\left(1-f\left(E\right)\right) + f\left(E\right)\left(ln\left(\dfrac{1 - f\left(E\right)}{f\left(E\right)}\right)\right)^{2}\biggr].
\label{rleco}
\end{align}
In Eq.~\ref{rleco}, the Fermi distribution is identified by $ f\left(E\right) $ and $ Li_{n}\left(z\right)= \sum_{k=1}^{\infty}z^{k}/k^{n} $ is the poly-logarithmic function. A quantitative assessment of $ L_{22}^{xy} $ or the anomalous thermal Hall conductivity can be carried out by evaluating the integral in Eq.~\ref{rleco}; a first step though lies in a reasonable estimation of $ \Omega\left(k\right)$ through an evaluation of $ \Delta $ in the \textit{off}-resonant condition. This is the program for the following section.

\vspace{0.3cm}
\section{Results}
\begin{figure}[!t]
\includegraphics[scale=0.7]{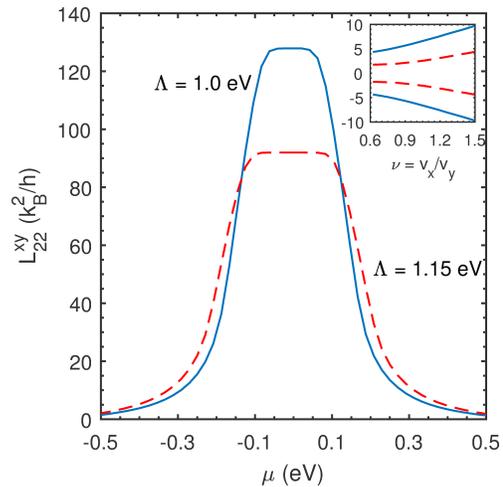}
\vspace{-0.31cm}
\caption{The numerically calculated anomalous thermal Hall conductivity $\left(L_{22}^{xy}\right)$ is shown here for a series of Fermi levels and a pair of $ \Lambda $ values indicated on the plot. The incident light is assumed to be right-circularly polarized. The inset traces the variation (monotonically increasing) of $ L_{22}^{xy} $ for several anisotropy $\left(\nu\right)$ ratios; the solid (broken) line is for $ \mu = 0.3 \left(0.4\right)\, eV $. For a left-circularly polarized light, the $ L_{22}^{xy} $ curve located under the origin is a mirror reflection of its upper right-circular counterpart for an identical $ \mu $.}
\vspace{-0.5cm}
\label{l22xy}
\end{figure}

\vspace{0.3cm}
We begin by setting the energy of the incident beam to a constant $ \hbar\omega = 8.0\,eV $ - the denominator of Eq.~\ref{commsf}. The term $ \Lambda^{2} = e^{2}A^{2}v_{x}v_{y} $ in the numerator is an adjustable variable assigned a pair of values (in eV): $ \Lambda = \lbrace 1.0, 1.15\rbrace $. The corresponding band gaps in $\mathrm{meV}$ using Eq.~\ref{commsf} are 125 and 165 respectively. To carry out a numerical integration of Eq.~\ref{rleco}, we start by assigning a $ k $-space cut-off radius of $ \vert k \vert = 0.25\,\AA^{-1} $ and obtain $ L_{22}^{xy} $ (Fig.~\ref{l22xy}, in unit of $ k_{B}^{2}/h $) for the selected pair of $ \Lambda $ and a series of Fermi levels $\left(\mu\right)$. The temperature is $ T = 100\, K $. We immediately recognize two key aspects of the plot: 1) The coefficient  $ L_{22}^{xy} $ is higher for a reduced value of $ \Lambda $ and 2) drops when $ \mu $ is deep into the valence or conduction band $\left(\mu < -\Delta ; \mu > \Delta\right) $; a rise, however, is observed for the condition $ -\Delta < \mu < \Delta $. To understand this behaviour, simply observe that $ L_{22}^{xy} $ is computed by estimating the integral in Eq.~\ref{rleco} for the set of occupied conduction and valence bands that are distinguished, inter alia, by opposite values of $ \Omega\left(k\right) $, the driver of anomalous heat flow. For $ \mu $ located in the band gap region, the contribution of the valence states to the overall integral is significantly higher vis-`a-vis its conduction counterparts reflected in a large value of $ L_{22}^{xy} $; contrarily for $ \mu $ positioned either in the conduction or valence region, the contributions are similarly matched but with a difference in sign leading to a cancellation and the observed low values. As representative numbers that account for the aforementioned cases, we set $ \mu $ to $ \lbrace 0,0.5\rbrace\, eV $ and selecting $ \Lambda = 1.0\,eV $, $ L_{22}^{xy} $ is approximately $ \lbrace 120, 1.2\rbrace \times k_{B}^{2}/h $, roughly two-orders of difference between them. A lowering of this reduction, however, is possible by tuning the anisotropy parameter, $ \nu = v_{x}/v_{y} $; indeed, for $ \mu = 0.3 \left(0.4\right)\,eV $, $ L_{22}^{xy} $ marked by a solid (broken) curve in the inset in Fig.~\ref{l22xy} rises monotonically for a greater degree of anisotropy. A straightforward explanation can be offered by noting the corresponding change in $\Omega\left(k\right)$ with anisotropy (also see, caption, Fig.~\ref{berth}). It is then reasonable to assume that such alterations to $ \nu $ can be achieved through variations of the slope $\left(v_{i} = \hbar^{-1}\partial E/\partial k_{i}\right)$ of the linear dispersion bands; a simple technique being the application of uniaxial or shear strain. The inset also shows a pair of downward sloping curves below the origin and symmetric with those above them; these are simply obtained by flipping the polarization of the incident illumination from right- to left-circular, the consequence of which physically manifests as a reversal in the direction of the heat current. It is pertinent to remark here that such complete reversals of heat flow leads to conceivable applications that exploit the phenomenon of thermal rectification, a preferential direction of heat transport~\cite{arora2017thermal}, particularly desirable in thermal transistors and logic circuits.

\vspace{0.3cm}
\subsection{Efficiency cycle}
\vspace{0.3cm}
In the preceding sections we have sought to quantitatively estimate the thermal Hall conductivity; it would also be useful to gauge and uncover methods (qualitatively) to better the efficiency of the thermal process which supplies a measure of the `anomalous' transverse current $\left(J_{xy}\right)$ per the `normal' longitudinal component $\left(J_{xx}\right)$. We can, employing Fourier's heat law, $ J = -L_{22}\nabla T $, write the efficiency as $ \Theta_{H} = J_{xy}/J_{xx} = L_{22}^{xy}/L_{22}^{xx} $. Note that $ \Theta_{H} $ is also the thermal Hall angle for the anomalous case. To this end, while we have an analytic representation of $ L_{22}^{xy} $ from Eq.~\ref{rleco}, $ L_{22}^{xx} $ can be assessed by a direct application of the Weidemann-Franz law (WFL).~\cite{chester1961law} We have, $ L_{22}^{xx} = \sigma_{xx}\mathcal{L}T $, where $ \sigma_{xx} $ is the electric conductivity and $ \mathcal{L} $ is the Lorenz number. The $ \sigma_{xx} $ can be crudely written as $ 0.5*e^{2}v_{x}^{2}\tau\mathcal{D}\left(\epsilon\right)$. The density of states is $ \mathcal{D}\left(\epsilon\right) = \sum_{k}\delta\left(\epsilon - E\left(k\right)\right) $ and the scattering time as a parameter is assumed to be constant in the range of energies around the Fermi level. The $ \mathcal{D}\left(\epsilon\right) $ is obtained from the dispersion relation of Eq.~\ref{diseq} and the identity $ \delta\left(\epsilon - E\left(k\right)\right) = \delta\left(k - k_{i}\right)/\vert g^{'}\left(k\right)\vert $, where $ k_{i} $ is a root of the equation $ g\left(k\right) = \epsilon - E\left(k\right) = 0 $. The superscript prime denotes a derivative. Tacitly, when WFL holds, the per spin $ L_{22}^{xx} $ at each DC (a pair exists) is
\begin{equation}
L_{22}^{xx} = \dfrac{\mathcal{L}T\Upsilon\epsilon}{2\pi\hbar^{2}}\int_{0}^{2\pi}d\theta\dfrac{1}{\left(v_{t}\sin\theta + v_{y}\sqrt{\nu^{2}\cos^{2}\theta + \sin^{2}\theta}\right)^{2}}.
\label{wfl}
\end{equation}
Here, $ \Upsilon = e^{2}v_{x}^{2}\tau $. This equation is integrable in a closed form for specific material parameters; also, note that we have ignored changes to conductivity arising from a photo-induced band gap. A quick inspection shows how the $ \nu $ parameter can be again brought to bear significant influence on $ L_{22}^{xx} $, and thereupon on the ratio, $ \Theta_{H} =  L_{22}^{xy}/ L_{22}^{xx} $. Succinctly, a complete account of optimizing $ \Theta_{H} $ involves an elaboration of its atomic and band structure under strain~\cite{cheng2017anisotropic} that impacts $ \nu $; this also rearranges the phonon dispersion and mobility curves in addition to a re-evaluation of $ \tau $ combining electron transport and deformation potential theory. 

\vspace{0.3cm}
\section{Final Remarks}
\vspace{0.3cm}
In closing, we set up an analytic model for the anisotropy-dependent anomalous thermal Hall conductance of a $ \beta $-borophene sheet placed under a circularly-polarized illuminating beam. The utility of such calculations lie in suitably harnessing the topology of the Fermi surface as a means to control thermoelectric behaviour. In case of $ \beta $-borophene, the promise of such control acquires significance in light of its underlying anisotropic thermal transport. On this note, surface hydrogenation of 2-\textit{Pmmn} borophene (an allotrope sharing the same space group of 59 with 8-\textit{Pmmn} borophene but with two atoms per unit cell) showed remarkable changes~\cite{nakhaee2018tight} to its structure; a first-principles calculation shows an elongation of the `a' lattice constant (along the \textit{x}-axis) by approximately $ 18\% $ over its pristine counterpart, thereby stretching the \textit{B-B} bond and re-calibrating the elastic force constants. The distortion to bond length and a newer set of phonon curves therefore further accentuates the anisotropy reflected in tangible reworking of the electric and thermal properties. Such atom-wise explorations that let us tap into the virtues of anisotropy remain to be seen for the 8-atom case discussed here. Besides, a more comprehensive description of the interplay between electron and drag-like phonon effects that may stand altered under structural microscopic adjustments may bring to light greater intricacies than understood heretofore.
 

\end{document}